\documentclass[prl,aps,superscriptaddress,preprint,floatfix,nofootinbib]{revtex4-1}

\usepackage{epsf}
\usepackage{graphicx}
\usepackage{sidecap}

\usepackage{color}

\def \beq {\begin{equation}}
\def \eeq {\end{equation}}
\begin{document}

\title{{Surface electronic structure of the topological Kondo insulator candidate correlated electron system SmB$_6$}}

%Surface electronic structure of a topological Kondo insulator candidate SmB$_6$

\author{M.~Neupane}
\affiliation {Joseph Henry Laboratory and Department of Physics,
Princeton University, Princeton, New Jersey 08544, USA}

\author{N.~Alidoust}\affiliation {Joseph Henry Laboratory and Department of Physics, Princeton University, Princeton, New Jersey 08544, USA}

\author{S.-Y.~Xu}
\affiliation {Joseph Henry Laboratory and Department of Physics, Princeton University, Princeton, New Jersey 08544, USA}

\author{T.~Kondo}
\affiliation {ISSP, University of Tokyo, Kashiwa, Chiba 277-8581, Japan}

\author{Y. Ishida}
\affiliation {ISSP, University of Tokyo, Kashiwa, Chiba 277-8581, Japan}

\author{D.J. Kim}
\affiliation {Department of Physics and Astronomy, University of California at Irvine, Irvine, CA 92697, USA}

\author{Chang~Liu}
\affiliation {Joseph Henry Laboratory and Department of Physics,
Princeton University, Princeton, New Jersey 08544, USA}

\author{I.~Belopolski}
\affiliation {Joseph Henry Laboratory and Department of Physics, Princeton University, Princeton, New Jersey 08544, USA}

\author{Y. J. Jo}
\affiliation{Department of Physics, Kyungpook National University, Daegu 702-701, Korea}

\author{T.-R. Chang}
\affiliation{Department of Physics, National Tsing Hua University, Hsinchu 30013, Taiwan}

\author{H.-T. Jeng}
\affiliation{Department of Physics, National Tsing Hua University, Hsinchu 30013, Taiwan}
\affiliation{Institute of Physics, Academia Sinica, Taipei 11529, Taiwan}

\author{T.~Durakiewicz}
\affiliation {Condensed Matter and Magnet Science Group, Los Alamos National Laboratory, Los Alamos, NM 87545, USA}

\author{L. Balicas}
\affiliation {National High Magnetic Field Laboratory, Florida State University, Tallahassee, Florida 32310, USA}
\author{H.~Lin}
\affiliation {Department of Physics, Northeastern University,
Boston, Massachusetts 02115, USA}
\author{A.~Bansil}
\affiliation {Department of Physics, Northeastern University,
Boston, Massachusetts 02115, USA}
\author{S.~Shin}
\affiliation {ISSP, University of Tokyo, Kashiwa, Chiba 277-8581, Japan}
\author{Z. Fisk}
\affiliation {Department of Physics and Astronomy, University of California at Irvine, Irvine, CA 92697, USA}
\author{M.~Z.~Hasan}
\affiliation {Joseph Henry Laboratory and Department of Physics,
Princeton University, Princeton, New Jersey 08544, USA}
\affiliation {Princeton Center for Complex Materials, Princeton University, Princeton, New Jersey 08544, USA}

\date{18 June, 2013}
\begin{abstract}

%\textbf{The Kondo insulator SmB$_6$ has long been known to exhibit low temperature transport anomalies whose origin is of great recent interest. By combining laser- and synchrotron-based angle-resolved photoemission techniques we uniquely access the surface electronic structure of the anomalous transport regime. We observe clear in-gap states (up to $\sim$ 4 meV) that are strongly temperature dependent which disappear in approaching the hybridization scale of 30 K supporting the theoretical prediction that the identified states are contingent upon the Kondo gap formation. Additionally, our observed Fermi surface odd-ness tied with the Kramers' point topology of the in-gap states, their co-existence with the 2D transport anomaly and more directly with the Kondo hybridization regime, as well as their robustness against thermal recycling taken together reported here for the first time collectively provide by-far the strongest case for a Z$_2$ topological Kondo phase in SmB$_6$. The laser and synchrotron based band structure study and Fermi surface topology provide the fundamental electronic parameters for the anomalous (true) groundstate of SmB$_6$ as well, which will be invaluably useful in designing, fabricating and interpreting functionalities in devices based on this exotic material.}

\textbf{The Kondo insulator SmB$_6$ has long been known to exhibit low temperature transport anomalies whose origin is of great interest. 
Here we uniquely access the surface electronic structure of the anomalous transport regime by combining state-of-the-art laser- and synchrotron-based angle-resolved photoemission techniques.
%By combining state-of-the-art laser- and synchrotron-based angle-resolved photoemission techniques we uniquely access the surface electronic structure of the anomalous transport regime. 
We observe clear in-gap states (up to $\sim$ 4 meV), whose temperature dependence is contingent upon the Kondo gap formation. In addition, our observed in-gap Fermi surface oddness tied with the Kramers' points topology, their coexistence with the two-dimensional transport anomaly in the Kondo hybridization regime, as well as their robustness against thermal recycling, taken together, collectively provide by-far the strongest evidence for protected surface metallicity with a Fermi surface whose topology is consistent with the theoretically predicted topological surface Fermi surface. Our observations of systematic surface electronic structure provide the fundamental electronic parameters for the anomalous Kondo ground state of the correlated electron material SmB$_6$.}
%, which will be invaluably useful in designing and fabricating functional devices based on this exotic material.}

%, at which the Kondo gap Kondo hybridized states exhibit an energy gap of 17 meV at momemtum (k) $\sim$ 0.3 \AA$^{-1}$ along $\Gamma$-\textbf{X} and remarkably feature clear low-lying states within a 4 meV window of the Fermi level which is well within the 17 meV energy gap      Moreover, our experimental results taken alone highlight an important low-energy component of the true groundstate of SmB$_6$  directly relevant for its anomalous charge and Hall transport behavior.

%irrespective of any topological theory.}

\end{abstract}
\date{\today}
\maketitle

%Introductions:

%These states disappear as the temperature is raised above 15 K in correspondence with the complete disappearance of 2D conductivity channels in SmB$_6$. While the topological nature of the in-gap metallic states cannot be ascertained without spin (spin-texture) measurements,

Materials with strong electron correlations often exhibit exotic ground states such as the heavy fermion behavior, Mott or Kondo insulation and unconventional superconductivity. Kondo insulators are mostly realized in the rare-Earth-based compounds featuring \textit{f}-electron degrees of freedom, which behave like a correlated metal at high temperatures, whereas a bulk bandgap opens at low temperatures through the hybridization \cite{Fisk, Phil, Riseborough} of nearly localized-flat \textit{f} bands with the \textit{d}-derived dispersive conduction band. With the advent of topological insulators \cite{Hasan, Hsieh, SCZhang, FuKaneMele, Xia, Hasan2} the compound SmB$_6$, often categorized as a heavy-fermion semiconductor \cite{Fisk, Phil, Riseborough}, attracted much attention due to the proposal that it may possibly host a topological Kondo insulator (TKI) phase at low temperatures where transport is anomalous \cite{Dzero,Takimoto, Dai}. The anomalous residual conductivity is believed to be associated with electronic states that lie within the Kondo gap \cite{Kimura,Menth, Allen, Cooley, Nanba, Nyhus, Alekseev, point_cont, Miyazaki, Denlinger}. To this date no angle-resolved photoemission spectroscopy (ARPES) has been used to gain insights on the surface electronic structure of this compound in this anomalous transport regime with appropriately matched required energy resolution and \textit{d/f} orbital contrast selectivity. Following the prediction of a TKI phase, there have been several transport measurements which include observation of a three-dimensional (3D) to two-dimensional (2D) crossover of the transport carriers below $T\sim 7$ K (which fully saturates around 5 K) and clear signatures of a coherent Kondo lattice hybridization onset around 30 K \cite{Fisk_discovery, Hall, tunelling}, suggesting three regimes of the transport behavior \cite{Hall, tunelling}. First, weakly interacting and single-ion resonance regime covering 30-150 K or higher; second, a 3D transport regime in the Kondo lattice hybridization regime covering a temperature range of 8-30 K; third, a 2D anomalous transport regime (6 K or lower) which is the ground state ($T \sim 0$) of this compound. However, to this date, no temperature dependent study of the low-energy states which are predicted to lie within 5 meV exist that provide the momentum-resolved nature of the low-lying bands and their Fermi surface character in this system. Despite many transport results available so far, a number of issues critically important for determining the relevance to the Z$_2$ topological band-theory of the ground state band-structure are not known. These include a direct knowledge of the following: first, topology or connectivity of the Fermi surface that is responsible for the apparently highly conducting 2D transport \cite{Fisk_discovery, Hall, tunelling}; second, the in-gap states that give rise to the Fermi surface must be temperature dependent and vanish below the coherent Kondo lattice hybridization scale of 30 K \cite{tunelling} if they are related to the topology; third, for the observed 2D transport to be of topological origin, it must be due to odd number of pockets which is not known from transport; fourth, Fermi pockets must enclose only an odd number of Kramers' points of the Brillouin zone lattice that relate to the bulk band inversion but not any other high symmetry point; and fifth, that the surface should exhibit non-trivial Berry's phase as a consequence \cite{FuKaneMele} of the facts mentioned in first to fourth, since these conditions are the TKI equivalents of the topological band insulators \cite{Dzero,Takimoto,Dai, FuKaneMele} (also see Supplementary Discussion and Supplementary Figures S1-S5). Since there are three distinct transport regimes, the true ground state lies only at the anomalous transport regime. Previous ARPES studies were not only limited to temperatures outside this interesting regime (typically higher than 10-20 K)\cite{Miyazaki, Denlinger} but also missed to identify features within 5 meV and their k-space maps due to a limited combined resolution, namely, the condition of \textquotedblleft{better than 5 meV and 7 K temperature combination}". By combining high-resolution laser- and synchrotron-based angle-resolved photoemission techniques, we uniquely access the surface electronic structure on the samples (growth batch) where transport-anomalies were identified \cite{Fisk_discovery, Hall}. We identify the in-gap states that are strongly temperature dependent and disappear before approaching the coherent Kondo hybridization scale. Our Fermi surface mapping covering the low-energy part of the in-gap states only, yet having the sample lie within the transport anomaly regime reveals an \textit{odd} number of pockets that enclose three out of the four Kramers' points of the surface Brillouin zone. This is remarkably consistent with the theoretical prediction for a topological surface Fermi surface in SmB$_6$ by Lu \textit{et al} \cite{Dai}. Our observed Fermi surface oddness, Kramers' point winding-only topology of the in-gap states, their direct correlation with the 2D transport anomaly and their coexistence with the robust Kondo lattice hybridization, as well as their robustness against thermal recycling all taken together by far provide the strongest evidence for the surface metallicity with a Fermi surface whose topology is consistent with the theoretically predicted topological surface Fermi surface of a Kondo insulator. The laser- and synchrotron-based bulk and surface band identifications of the low temperature phase by themselves collectively provide the fundamental quantitative electronic parameters for the anomalous ground state bandstructure of SmB$_6$.
% which will be invaluably useful for designing, fabricating and interpreting new devices and functionalities made out of this exotic material in general.

\bigskip
\bigskip
\textbf{Results}

\textbf{Crystal structure and transport anomaly}

SmB$_6$ crystallizes in the CsCl-type structure with the Sm ions and the B$_6$ octahedra being located at the corner and at the body center of the cubic lattice, respectively (see Fig. 1a). The bulk Brillouin zone (BZ) is a cube made up of six square faces. The center of the cube is the $\Gamma$ point, whereas the centers of the square faces are the $X$ points. Because of the inversion symmetry of the crystal, each $X$ point and its diametrically opposite partner are completely equivalent. Therefore, there exist three distinct $X$ points in the BZ, labeled as $X_1$, $X_2$ and $X_3$. It is well established that the low-energy physics in SmB$_6$ is constituted of the non-dispersive Sm $4f$ band and the dispersive Sm $5d$ band located near the $X$ points \cite{Fisk_discovery, Hall, tunelling, Miyazaki, Denlinger}. To cross-check the established properties, we present the temperature dependent resistivity profile as well as the overall electronic band-structure (Figs. 1c-e) for samples used in our ARPES measurements. The resistivity profile shows a rise for temperatures below 30 K, which is in agreement with the opening of the Kondo gap at the chemical potential. Thus, the Fermi level in our sample lies within the bandgap and sample is bulk insulating. Moreover, at $T\leq$ 6 K, the resistivity starts/begins to saturate (which fully saturates below 5 K, Fig. 1), indicating the onset of 2D transport anomaly \cite{Menth, Cooley}.

\bigskip
\bigskip

\textbf{Bulk band structure}

In Figs. 1d and 1e, we present ARPES intensity profiles over a wide binding energy scale measured with a synchrotron-based ARPES system using a photon energy of 26 eV. The non-dispersive Sm 4$f$ states near the Fermi level are seen and clearly identified in the integrated energy distribution curves (EDCs). The location of the flat bands are estimated to be at binding energies of 15 meV, 150 meV and 1 eV, which correspond to the $^6$H$_{5/2}$, $^6$H$_{7/2}$ and $^6$F multiplets of the Sm$^{2+}$ ($f^6$ to $f^5$) final states, respectively, in agreement with the earlier reports \cite{Miyazaki, Denlinger} (see also Supplementary Discussion and Supplementary Figure S6). The dispersive features we observe originate from the Sm $5d$-derived bands and a hybridization between the Sm $5d$ band and Sm 4$f$ flat band is visible especially around 150 meV binding energies confirming the Kondo ingredients of the electronic system in our study (Figs. 1d and 1e).

\bigskip
\bigskip
\textbf{In-gap states and Fermi surface}

To search for the predicted in-gap states within 5 meV of the Fermi level, we employ a laser-based ARPES system providing $\Delta{E}\sim4$ meV coupled with a low temperature ($T\simeq5$ K) capability. This combination allows us to study the transport anomaly regime with appropriate energy resolution. Furthermore, the choice of 7 eV photons from a laser source is due to the fact that it allows us to improve the relative photoionization cross-section for the \textit{d/f} content of the hybridized band. At 7 eV, the cross section for the $f$ orbitals is weaker than it is at typical synchrotron photon energies \cite{Crosssection}, so the measured states have better correspondence to the partial $d$-orbital character of the hybridized band which is favorable for isolating the topological surface states \cite{Dzero,Takimoto, Dai}. This is further important for the identification of the in-gap states at all Kramers' points and systematically probing their simultaneous or coupled temperature evolution (if any) as well as measuring their k-space maps. At higher photon frequencies as in the case of synchrotron-based ARPES detections, the \textit{f}-electron contribution can be quite large, thus masking out the low-energy 2D states due to the \textit{f}-component of the strong band tails. The current study thus focuses on testing the proposal of k-space maps of in-gap states and their connection with the 2D transport regime by optimizing the best possible scenarios along the line of specific requirements of the theoretical predictions in refs. \cite{Dzero,Takimoto, Dai}.

Since the low-energy physics including the Kondo hybridization process occurs near the three $X$ points (Fig. 2a) in the bulk BZ and the $X$ points project onto the $\bar{X}_1$, $\bar{X}_2$, and the $\bar{\Gamma}$ points at (001) surface (Fig. 1b), the Kramers' points of this lattice are $\bar{X}_1$, $\bar{X}_2$, $\bar{\Gamma}$ and $\bar{M}$ and we need to systematically study the connectivity (winding) of the in-gap states around these points. In Fig. 2c, we show our measured ARPES spectral intensity integrated in a narrow ($\pm0.15$ $\textrm{\AA}^{-1}$) momentum window and their temperature evolution around the $\bar{X}$ point. At temperatures above the hybridization scale, only one spectral intensity feature is observed around $E_{\mathrm{B}}\sim12$ meV in the ARPES EDC profile. As temperature decreases below 30 K, this feature is found to move to deeper binding energies away from the chemical potential, consistent with the opening of the Kondo hybridization gap while Fermi level is in the insulating gap (bulk is insulating, according to transport, so Fermi level must lie in-gap at 6 K). At lower temperatures, the gap value of hybridized states at this momentum space regime is estimated to be about 16 meV. More importantly, at a low temperature $T\simeq6$ K corresponding to the 2D transport regime, a second spectral intensity feature is observed at the binding energy of $E_{\mathrm{B}}\sim4$ meV, which lies inside the insulating gap. Our data thus experimentally show the existence of in-gap states. Remarkably, the in-gap state feature is most pronounced at low temperature $T\simeq 6$ K in the 2D transport regime, but becomes suppressed and eventually vanishes as temperature is raised before reaching the onset for the Kondo lattice hybridization at 30 K. The in-gap states are found to be robust against thermal cycling, since lowering the temperature back down to 6 K results in the similar spectra with the re-appearance of the in-gap state features (Re\_6K in Fig. 2c). Within a 30-hour period we did not observe any significant aging effect in our samples while maintaining an ultra-high vacuum level better than 4$\times$10$^{-11}$ Torr. The observed robustness against thermal recyclings counts against the possibility of non-robust (trivial) or non-reproducible surface states. We further performed similar measurements of low-lying states focusing near the $\bar{\Gamma}$ point (projection of the $X_3$) as shown in Fig. 2d. Similar spectra reveal in-gap state features prominently around $E_{\mathrm{B}}\sim3-4$ meV at $T\simeq6$ K, which clearly lie within the Kondo gap and exhibit similar (coupled) temperature evolution as seen in the spectra obtained near the $\bar{X}$ point. This feature also disappears before reaching the Kondo hybridization scale which is non-suggestive of  the possibility that this is the bottom of the conduction band. 
The fact that our sample is bulk insulating suggests that the Fermi level is not at the bottom of the conduction band, which would make the sample highly metallic in transport.
However, the surface Fermi level might also lie within the conduction band while it is bulk insulating which is another form of metallicity discussed in ref. \cite{HsiehNat}.

%Since our sample is bulk insulating, surface Fermi level is at the bottom of the band which would make the sample hugely metallic.

On experimentally establishing the existence of in-gap states, we study their momentum-resolved structure or the k-space map for investigations regarding their topology: (1) the number of surface state pockets that lie within the Kondo gap; (2) the momentum space locations of the pockets (whether enclosing or winding the Kramers' points or not). Fig. 3a shows a Fermi surface map measured by setting the energy window to cover $E_{\mathrm{F}}\pm4$ meV, which ensures the inclusion of the in-gap states (that show temperature dependence consistent with coupling to the Kondo hybridization) within the Fermi surface mapping data as identified in Figs. 2c and 2d, at a temperature of 6 K inside the 2D transport anomaly regime under the \textquotedblleft{better than 5 meV and 7 K combined resolution condition}". Our Fermi surface mapping reveals multiple pockets which consist of an oval-shaped as well as nearly circular-shaped pockets around the $\bar{X}$ and $\bar{\Gamma}$ points, respectively. No pocket was seen around the $\bar{M}$-point which was measured in a synchrotron ARPES setting. Therefore, the laser ARPES data capture all the pockets that exist while the bulk is insulating. This result is striking by itself from the point of view that while we know from transport that the bulk is insulating, ARPES shows large Fermi surface pockets (metallicity of the surface) at this temperature. Another unusual aspect is that not all Kramers' points are enclosed by the in-gap states. It is known that all trivial surface states must always come with even number of pockets, which include surface states that are due to reconstruction or surface polarity-driven or surface impurity-driven since there is no way to get around the fermion doubling problem \cite{FuKaneMele}. Trivial surface states in non-magnetic crystals must always come in pairs, otherwise a cornerstone of quantum mechanics, Kramers' theorem would be violated \cite{FuKaneMele}. Another rigorous fact also mandated by the quantum mechanics on a lattice on general grounds is that for an inversion symmetric crystal (which is the case for SmB$_6$) harboring spin-orbit coupling large enough to invert the bulk bands (as in SmB$_6$ according to all bulk band calculations), if Kramers' theorem holds, the surface states cannot be degenerate at non-high-symmetry points even if they are trivial (see, \cite{FuKaneMele, Hasan, SCZhang, Hasan2}). These conditions hold true irrespective of the chemical details or surface reconstruction or polarity-driven or dangling-bond origin nature of the surface states, and allow us to count the Fermi surface pockets based on their momentum-space winding in a Mod 2 count \cite{FuKaneMele, Hasan, SCZhang, Hasan2}. Our observed Fermi surface thus consists of 3 (or odd number Mod 2 around each Kramers' point) pockets per Brilluoin zone and each of them wind around a Kramers' point only and this number is odd (at least three within our resolution). Therefore, our measured in-(Kondo)gap states lead to a very specific form of the Fermi surface topology (Fig. 3) that is remarkably consistent with the theoretically predicted topological surface state Fermi surface expected in the TKI ground state phase despite the broad nature of the contours. We further present the measured energy-momentum cuts at $6$ K (Figs. 3c and d), where low-energy states consistent with the observed surface Fermi surface topology and their odd numbered crossing behavior (Fig. 3a) are better identified than the k-space maps themselves (see Supplementary Discussions and Supplementary Figures S1-S5). The broadness is due to extremely small Fermi velocity especially near the zone center. The Fermi velocity of the low-lying states is estimated to be less than $\simeq0.3$ eV$\cdot\mathrm{\AA}$ and is consistent with the data, however, the details of dispersion within the Kondo gap are not well-resolved, which is due to the finite quasi-particle lifetime broadening inevitable in a small Kondo gap material. This intrinsic life-time, an expected effect, is elaborated and quantified with a numerical simulation in the Supplementary Discussion and Supplementary Figure S2. Such broadening cannot be eliminated (not a resolution or data quality issue) as long as Kondo gap is on the order of 15-20 meV in a correlated material where sample mobility is on the order of several thousands cm$^2 \cdot \mathrm{V^{-1}} \cdot \mathrm{sec}^{-1}$ which is the best possible scenario to this date.

%\bigskip
\bigskip
\textbf{Synchrotron-based ARPES measurements}

Since for the laser ARPES, the photon energy is fixed (7 eV) and the momentum window is rather limited (the momentum range is proportional to $\sqrt{h\nu-W}$, where $h\nu$ is the photon energy and $W\simeq4.5$ eV is the work function), we utilize synchrotron based ARPES measurements to study the low-lying state as a function of photon energy as demonstrated in Bi-based topological insulators \cite{Hasan, SCZhang, Hasan2}. Figs. 4a and b show the energy-momentum cuts measured with varying photon energies. Clear $E-k$ dispersions are observed (black dotted lines in Fig. 4b as a guide to the eye) within a narrow energy window near the Fermi level. The dispersion is found to be unchanged upon varying photon energy, supporting their quasi-2D nature (see Fig. 4c). The observed quasi-2D character of the signal within 10 meV where surface states reside does suggest consistency with the surface nature of the in-gap states. Because of the combined effects of energy resolution ($\Delta{E}\geq10$ meV, even though the sample temperature, 7 K, is near the anomalous transport regime) and the intrinsic self-energy broadening coupled with the higher weight of the \textit{f}-part of the cross-section and the strong band tails (see Supplementary Discussion), the in-gap states are intermixed with the higher energy bulk bands' tails. We speculate that the lack of $k_\mathrm{z}$ dispersion in the bands lying below 10 meV is due to some band-bending effect of the bulk bands that must occur in an intrinsic bulk insulator/semiconductor. A bent-bulk band near a surface should naturally exhibit weak k$_z$ dispersion since the effect is confined near the surface only. To isolate the in-gap states from the bulk band tails that have higher cross-section at synchrotron photon energies, it is necessary to have energy resolution (not just the low working temperature) better than half the Kondo gap scale which is about 7 meV or smaller in SmB$_6$. 
%These facts taken together further demonstrate the justification for the choice of a \textit{combination} of laser and synchrotron-based complementary ARPES study of this compound to prove the existence of the surface states.

\bigskip

\bigskip
\textbf{Discussion}

We now discuss the robust observations in our data and their connection to the theoretical prediction of the TKI phase. We have systematically studied the surface electronic structure of SmB$_6$ at the transport anomaly regime ($T\simeq6$ K), where the transport character of the conduction electrons are predominantly 2D. We observe low-lying features extending to the Fermi level which lie inside the Kondo insulating gap. As the temperature is raised across the transport anomaly (beyond $7-8$ K) and the Kondo lattice hybridization onset is approached, the in-gap states (main features) become suppressed before reaching 30 K, which suggests that the dominant in-gap states are contingent upon the existence of strong and saturating Kondo hybridization and the existence of a robust and saturated insulating bulk gap. The in-gap states are found to be robust and reproducible against thermal cycling of the sample in and out of the Kondo regime only, excluding the possibility of unrelated trivial surface states due to dangling bonds or polarity-driven states which can robustly appear at much higher temperatures due to surface chemistry. Polarity-driven (non-topological) surface states must also come in even number of pockets at any high symmetry point of the Brillouin zone unrelated to the Kramers' point topology. Furthermore, our k-space mapping covering the in-gap states shows distinct pockets that enclose three (not four) Kramers' points of the surface Brillouin zone, which are remarkably consistent with the theoretically predicted topological surface Fermi surface in the TKI ground state phase. Previous (and present) bulk band calculations have reconfirmed the strong role of spin-orbit coupling in the bulk bands in this inversion symmetric material which is suggestive of the surface states (trivial or non-trivial) on this compound being non-degenerate. This allows us to compare the surface states with theory that wind around the Kramers' points.
% to consist of three or higher odd numbered nondegenerate pockets, even though they are broad due to low V$_F$ and large intrinsic self-energy (see Supplementary Discussion and Supplementary Figure S2), which is the principle criterion for a Z$_2$ topological insulator in a Kondo setting.
In the presence of bulk spin-orbit coupling and band inversion, these surface states must carry 3$\pi$ Berry's phase which is equivalent to $\pi$ (3$\pi$ Mod 2$\pi$). According to ref. \cite{FuKaneMele}, the most important criterion for a topological insulator is that of possessing odd number (1, 3, 5, ...) of surface Fermi pockets of non-degenerate bands in a strong spin-orbit system with bulk band inversion and the rest follows. These theoretical criteria directly imply a \textit{net} surface Berry's phase of $\pi$ \cite{FuKaneMele}. Another independent way of proving a total 3$\pi$ Berry's phase in this system would be to carry out a direct spin-ARPES measurement as demonstrated previously in Bi-based compounds in our earlier works (see reviews \cite{Hasan, SCZhang, HsiehNat}). It is evident that the spin-resolved measurement in this system is currently not feasible to address the same issue. The energy resolution of spin-ARPES at synchrotrons is about 30-50 meV $\gg$ 5 meV in-gap state scale [see \cite{spinarpes, HsiehNat}], and with laser, 14 meV $\gg$ 5 meV [see \cite{laserspinarpes}] and simultaneously accessible lowest temperature in combination with best spin-resolution is about 20 K $\gg$ 7 K which is much larger than the 2D transport anomaly scale only within which surface states are unmixed with the bulk bands. Above 8 K, transport is 3D; therefore, bulk band must intermix in low-energies. Since the Kondo gap is small the intrinsic lifetime is large (see Supplementary Discussion) placing the system out of parameter range for all state-of-the-art spin-ARPES measurement conditions. The fact that the surface states disperse in such a narrow Kondo gap of 10-15 meV which makes it difficult to track them with spin-ARPES also suggests that these states are quite interesting from the view point of interaction effects and there does exist correlation effects in these surface states. This provides an exciting opportunity to study the interplay between Z$_2$ order and electron-electron correlation via interface effects in future experiments. One exciting future direction to pursue would be to grow this material epitaxially on a suitable superconductor, which then via the proximity effect would enable superconductivity in a strongly correlated setting. Such superconductivity is likely to be unconventional. Regardless of the future works, our observed odd Fermi surfaces of the in-gap states, their temperature dependence across the transport anomaly, as well as their robustness against thermal recycling, taken together collectively not only provide a unique insight illuminating on the nature of this 50-year-old puzzle in heavy-fermion physics, but also serve as a future guideline to investigate other novel heavy-fermion materials in connection to their exotic physics and transport anomalies.

\bigskip
\bigskip
\textbf{Methods}
\newline
\textbf{Electronic structure measurements.}
Synchrotron-based ARPES measurements of the electronic structure were performed at the Synchrotron Radiation Center, Wisconsin, Advanced Light Source (ALS), Berkeley, and Stanford Radiation Lightsource (SSRL), Stanford equipped with high efficiency R4000 electron analyzers. Separate, high-resolution and low temperature, low-photon-energy ARPES measurements were performed using a Scienta R4000 hemispherical analyzer with an ultraviolet laser ($h \nu = 6.994 $ eV) at the Institute for Solid State Physics (ISSP) at University of Tokyo. The energy resolution was set to be better than 4 meV and 10-20 meV for the measurements with the laser source and the synchrotron beamlines, respectively. The angular resolution was set to be better than 0.1$^{\circ}$ for the laser measurements and better than 0.2$^{\circ}$ for all synchrotron measurements.
% Further details regarding experimental procedures and data analysis are presented in the Supplementary Information (SI) section.
\newline

\textbf{Author contributions}
\newline
M.N., N.A., S.-Y.X., T.K., Y.I. performed the experiments with assistance from C.L., I.B., T.D., S.S.
and M.Z.H.; D.J.K., Z.F. provided samples that were characterized by Y.J.J., T.D. and L.B. and Z.F.;
T.R.C., H.T.J., H.L., A.B. carried out band calculations; M.Z.H. was responsible for the conception and the overall direction, planning and integration among different research units.

\textbf{Additional information}
\newline
Supplementary Information accompanies this paper at http://www.nature.com/naturecommunications
\newline
Competing financial interests: The authors declare no competing financial interests.
\newline
\*Correspondence and requests for materials should be addressed to
\newline
M.Z.H. (Email: mzhasan@princeton.edu).

\textbf{Competing financial interests:} The authors declare no competing financial interests.

\begin{figure*}
\centering
\includegraphics[width=16.5cm]{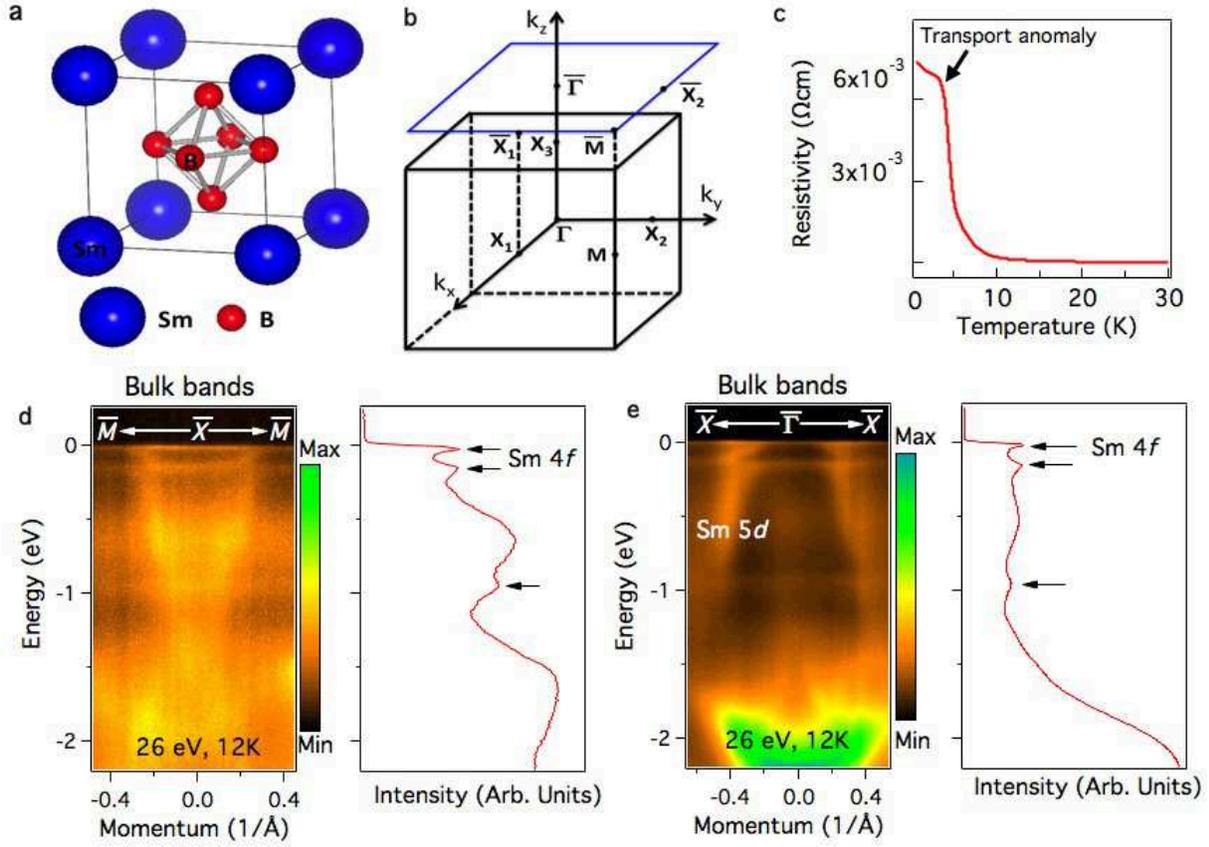}
\caption{\textbf{Brillouin zone symmetry and bulk band structure of SmB$_6$}. \textbf{a,} Crystal structure of SmB$_6$. Sm ions and B$_6$ octahedron are located at the corners and the center of the cubic lattice structure. \textbf{b,} The bulk and surface Brillouin zones of SmB$_6$. High-symmetry points are marked. \textbf{c,} Resistivity-temperature profile for samples used in ARPES measurements. \textbf{d, e,} Synchrotron-based ARPES dispersion maps along the ${\bar{M}}-{\bar{X}}-{\bar{M}}$ and the ${\bar{X}}-{\bar\Gamma}-{\bar{X}}$ momentum-space cut-directions. Dispersive Sm 5$d$ band and non-dispersive flat Sm 4$f$ bands are observed, confirming the key ingredient for a heavy fermion Kondo system. The momentum-integrated energy distribution curves are also plotted to clearly demonstrate the flat bands of Sm 4$f$. The surface BZ Kramers' points are $\bar\Gamma$, $\bar{X}_1=\bar{X}_2$, and $\bar{M}$.}
\end{figure*}

\begin{figure*}
\centering
\includegraphics[width=16.5cm]{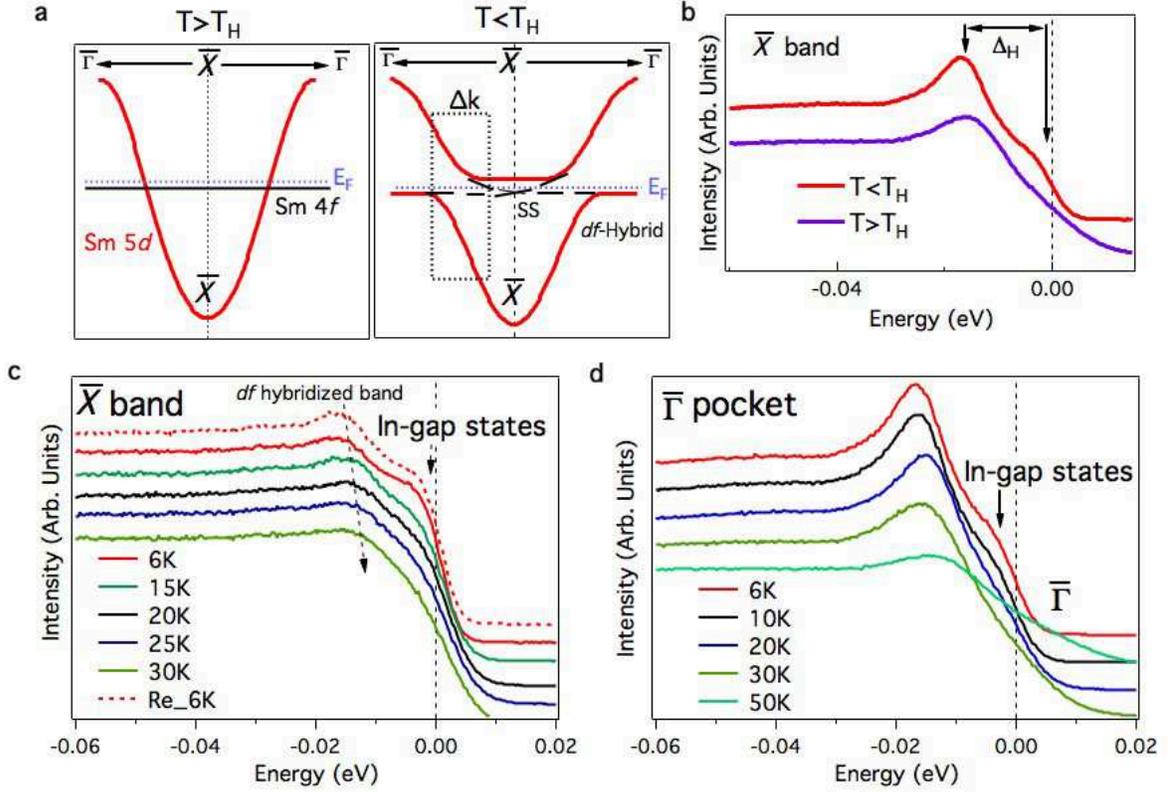}
\caption{\textbf{Identification of temperature-dependent in-gap states.} \textbf{a,} Cartoon sketch depicting the basics of Kondo lattice hybridization at temperatures above and below the hybridization gap opening. The blue dashed line represents the Fermi level in bulk insulating samples such as SmB$_6$ (since the bulk of the SmB$_6$ is insulating, the Fermi level must lie within the Kondo gap). The theoretically predicted topological surface states within the Kondo gap are also shown in this cartoon view (black dash lines) based on refs. \cite{Takimoto, Dai}. The black dash rectangle (Fig. 2\textbf{a}, right) shows the approximate momentum window of our laser-ARPES measurements between $k_1=0.1$ $\textrm{\AA}^{-1}$ to $k_2=0.4$ $\textrm{\AA}^{-1}$. \textbf{b,} Partially momentum-integrated ARPES spectral intensity in a $\pm0.15$ $\textrm{\AA}^{-1}$ window ($\Delta{k}$ defined in panel \textbf{a}) above and below the Kondo lattice hybridization temperature ($T_{\mathrm{H}}$). \textbf{c,} Momentum-integrated ARPES spectral intensity centered at the $\bar{X}$ point at various temperatures. \textbf{d,} Analogous measurements as in \textbf{c} but centered at the $\bar{\Gamma}$ pocket ($\Delta{k}=0.3$ $\textrm{\AA}^{-1}$). ARPES data taken on the sample after thermally recycling (6 K up to 50 K then back to 6 K) is shown by Re$\_$6K, which demonstrates that the in-gap states are robust and protected against thermal recycling.}
\end{figure*}

\begin{figure*}
\centering
\includegraphics[width=15.5cm]{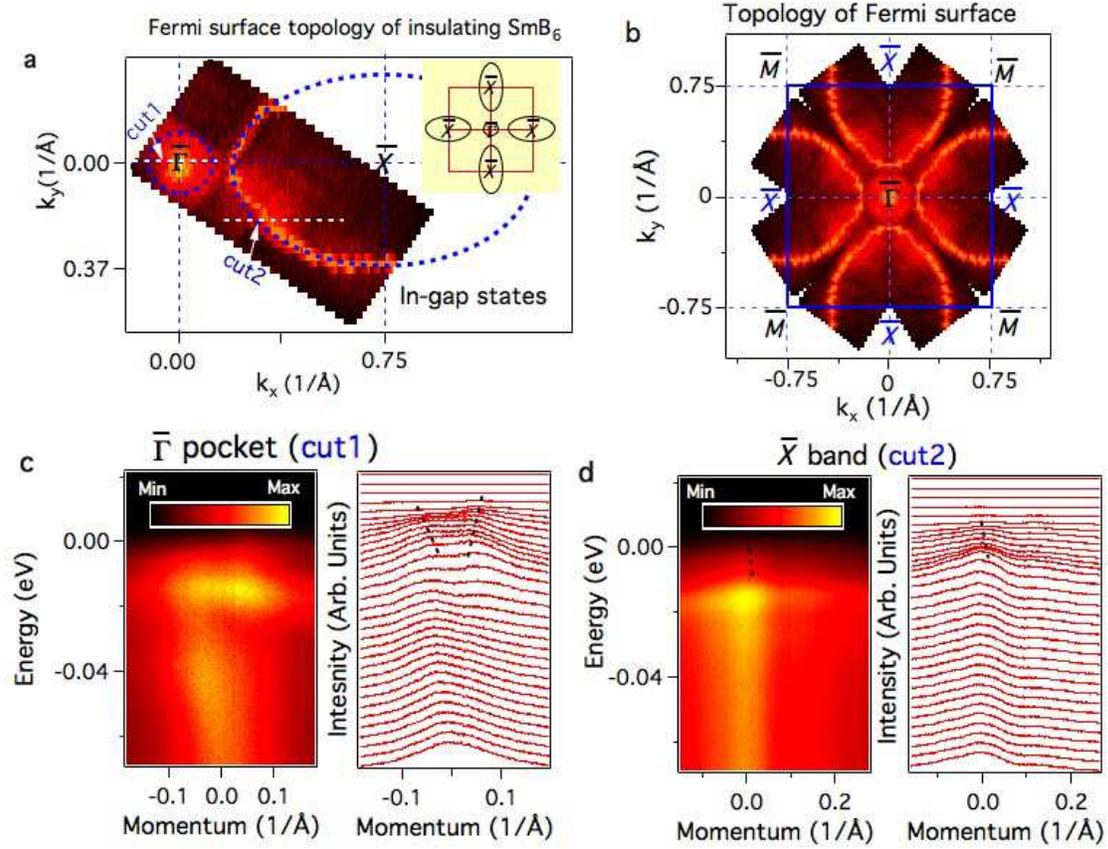}
\caption{\textbf{Topology of surface Fermi surface of bulk insulating SmB$_6$.} \textbf{a,} A Fermi surface map of bulk insulating SmB$_6$ using a 7 eV laser source at a sample temperature of $\simeq6$ K (resistivity$=5$ m$\Omega$cm), obtained within the $E_{\mathrm{F}}\pm4$ meV window, which captured all the low energy states between 0 to 4 meV binding energies, where in-gap surface state's spectral weight contribute most significantly within the insulating Kondo gap . Intensity contours around $\bar{\Gamma}$ and ${\bar{X}}$ reflect low-lying metallic states near the Fermi level, which is consistent with the theoretically predicted Fermi surface \textit{topology} of the topological surface states. The dashed lines around ${\bar{X}}$ and $\bar{\Gamma}$ points are guides to the eye. Inset shows a schematic of surface Fermi surface topology in the first Brillouin zone, where the pockets only enclose the Kramers' points required by the nontrivial Z$_2$ topological phase. \textbf{b,} Symmetrized surface Fermi surface plot based on the data shown in \textbf{a} without any background subtraction. \textbf{c,d,} Dispersion of the low-lying states (left) and the corresponding momentum distribution curves (right) near the $\bar{\Gamma}$ pocket (\textbf{c}) and the $\bar{X}$ pocket (\textbf{d}). Black dashed lines in \textbf{c} and \textbf{d} are guides to the eye, which track the significant dispersive features near the Fermi level. The intrinsically self-energy broadening of these low-lying states are theoretically presented in the Supplementary Discussion and Supplementary Figure S2. 
%The broadening is not limited by the laser-ARPES resolution but reflects the large self-energy intrinsic to a small Kondo gap material.
}
\end{figure*}

\begin{figure*}
\centering
\includegraphics[width=17.5cm]{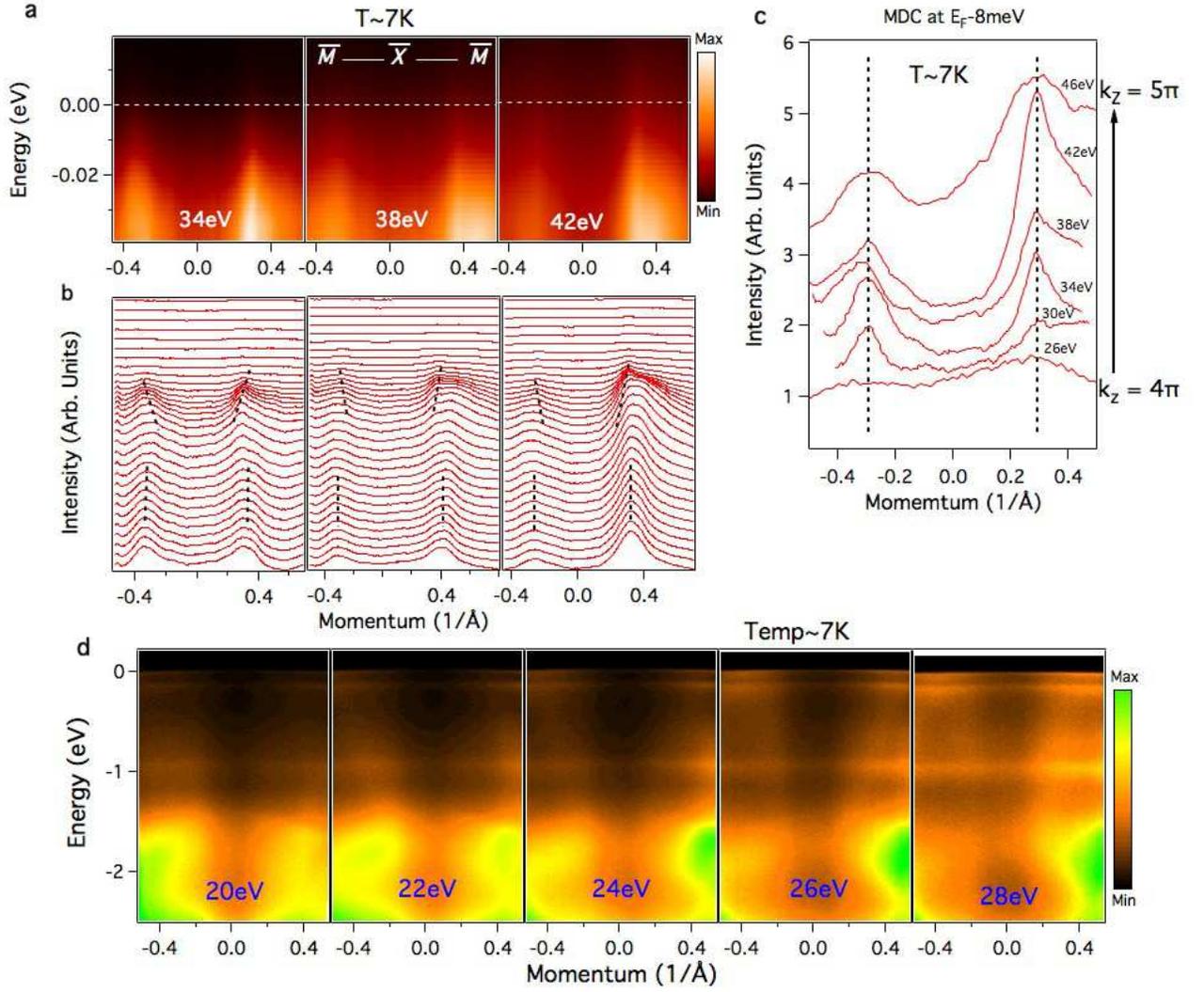}
\caption{\textbf{Synchrotron-based $k_z$ dispersion study of low-energy states at 7 K.} \textbf{a,} Synchrotron-based ARPES energy momentum dispersion maps measured using different photon energies along the ${\bar{M}}-{\bar{X}}-{\bar{M}}$ momentum space cut-direction. Incident photon energies used are noted on the plot. \textbf{b,} Momentum distribution curves (MDCs) of data shown in \textbf{a}. The peaks of the MDCs are marked by dashed lines near the Fermi level, which track the dispersion of the low-energy states. \textbf{c,} MDCs in the close vicinity of the Fermi level (covering the in-gap states near the gap edge) integrated within the energy window of [$E_{\mathrm{F}}$ - 8 meV, $E_{\mathrm{F}}$] are shown as a function of photon energy which covers the $k_z$ range of $4\pi$ to $5\pi$ at 7 K. \textbf{d,} ARPES dispersion maps measured along the ${\bar{\Gamma}}-{\bar{X}}-{\bar{\Gamma}}$ momentum space cut-direction over a wider energy range at various photon energies are also shown at $T=7$ K. Low-energy features are best resolved in data presented in \textbf{b} and \textbf{c}.}
\end{figure*}

%
%
%\begin{figure*}
%\centering
%\includegraphics[width=16.5cm]{Fig5}
%\caption{\textbf{Absence of bulk $\Gamma$-pocket and coupled temperature dependence of the in-gap states.}
%(\textbf{a}) Hybridized bands in the vicinity of the Fermi level along the $\Gamma-{\textrm{X}}$ momentum space cut. A bandgap of about 15 meV is obtained in our calculation (see supplementary information for details)
%(\textbf{b}) Orbital decomposed band structure near Fermi level along the $\Gamma-{\textrm{X}}$ line. The size of blue and yellow spheres are proportional to the weight of Sm 5$d$ and 4$f$ orbital, respectively.
%(\textbf{c}) A comparison of integrated EDCs featuring the $\Gamma$ pocket and the ${\textrm{X}}$ point bands and the low-lying in-gap states. }
%A gap value of about 14 meV is observed in both cases\end{figure*}

\end{document}